\documentclass[a4paper,11pt]{article}
\pdfoutput=1 % if your are submitting a pdflatex (i.e. if you have
\usepackage{jheppub} % for details on the use of the package, please
\usepackage[T1]{fontenc} % if needed
\usepackage{subcaption}
\usepackage{verbatim}
\usepackage{mathrsfs}
\usepackage{physics}
\usepackage{amsthm}
\usepackage{mathtools}
\usepackage{enumerate}

\newcommand{\del}{\partial}

\renewcommand{\hat}{\widehat}

\newcommand{\reals}{\mathbb{R}}
\newcommand{\comps}{\mathbb{C}}

\newcommand{\A}{\mathcal{A}}

\renewcommand{\H}{\mathcal{H}}

\theoremstyle{definition}

\numberwithin{theorem}{section}

\title{A classical model for semiclassical state-counting}
\author{Jonathan Sorce}
\abstract{In the type II von Neumann algebras that appear in semiclassical gravity, all states have infinite entropy, but entropy differences are uniquely defined. 
Akers and I have shown that the entropy difference of microcanonical states has a relative state-counting interpretation in terms of the additional (finite) number of degrees of freedom that are needed to represent the ``larger-entropy'' state supposing that one already has a representation of the ``smaller-entropy'' state, and supposing that one is restricted to act with gauge-invariant operators.
This short paper explains some of the curious features of relative state-counting by analogy to the classical limit of quantum statistical mechanics.
In this analogy the preferred family of renormalized traces becomes the preferred family of symplectic measures on phase space; the trace-index of infinite-dimensional subspaces becomes the ratio of phase space volumes; and the restriction that one must act with gauge-invariant operators becomes the restriction that one must act with symplectomorphisms.
Because in the phase-space analogy one has exact control over the quantum deformation away from the classical theory, one can see precisely how the relevant aspects of the classical structure are inherited from the quantum theory --- though even in this simple setting, it is a nontrivial technical task to show how classical symplectomorphisms emerge from the underlying quantum theory in the $\hbar \to 0$ limit.
}
\begin{document}
\maketitle

\section{Introduction}

Black holes are quantum systems; but unlike most quantum systems, they possess two different levels of quantumness.
This is because there are two independent parameters that affect the quantum behavior of black holes: Newton's constant $G_N,$ and Planck's constant $\hbar.$
When the surface area of a black hole is parametrically large relative to $G_N \hbar$, effects from quantum gravity are suppressed; but if there are matter fields present with an action of order $\hbar,$ then these fields must be treated quantumly.
The limit $G_N \hbar \to 0$ with $\hbar$ finite is called the \textit{semiclassical} limit of quantum gravity; quantum effects are still relevant, but only the ones due to matter fields.

Semiclassical black holes are typically assigned an entropy
\begin{equation} \label{eq:Bekenstein-Hawking}
	S_{\text{BH}} = \frac{\text{Area}_{\text{BH}}}{4 G_N \hbar},
\end{equation}
and there is very strong evidence\footnote{See for example \cite{Gibbons:entropy, Sen:1995in, Strominger:strings, Callan:micro, Horowitz:micro,Maldacena:micro, Johnson:micro, Strominger:CFT, Hartman:light-spectrum, Penington:replica, Iliesiu:2022kny, Balasubramanian:2022gmo}.} that in the semiclassical limit $G_N \hbar \to 0,$ a quantum black hole has an effective Hilbert space dimension that scales parametrically as the exponential of the semiclassical entropy.
There are, however, relatively few settings in which the microstructure of the black hole Hilbert space can be understood all the way from the quantum regime into the semiclassical one.
It is therefore productive to ask how much information about the fully quantum black hole Hilbert space can be understood from semiclassical considerations alone.
An obvious obstruction is that in the strict semiclassical limit, the black hole entropy $S_{\text{BH}}$ diverges.
Nevertheless, if one tries to compare two black holes with areas that differ from one another only at order $O(G_N \hbar)$, then the entropy \textit{difference} between these black holes is finite in the strict semiclassical limit, and stands a chance at being interpreted semiclassically.

In the past several years, a novel technology has been developed \cite{Witten:crossed-product, Chandrasekaran:deSitter, Chandrasekaran:microcanonical, Sorce:types, Jensen:general-subregions, Kudler-Flam:black-holes,Kudler-Flam:approximation, Akers:state-counting} for defining and computing entropy differences in the strict semiclassical limit, without putting in any details about the underlying quantum theory from which this limit emerged.
The point is that quantization of the linear-in-$G_N \hbar$ perturbations to certain fixed black hole backgrounds produces a Hilbert space $\H_{\text{s.c.}}$ carrying a von Neumann algebra $\A_{\text{s.c.}}$ that is a so-called type II factor.
The key feature of type II factors --- see for example \cite[section 6]{Sorce:types} --- is that while all of their operators have infinite trace, they possess a preferred family of renormalized traces for which there is a dense set of finite-trace operators.
A positive operator $\rho \in \A_{\text{s.c.}}$ with finite renormalized trace $\tau(\rho)$ is treated as a density matrix and is assigned the entropy
\begin{equation}
	S_{\tau}(\rho) = - \tau\left( \frac{\rho}{\tau(\rho)} \log \frac{\rho}{\tau(\rho)}\right).
\end{equation}
The choice of renormalized trace is ambiguous only up to a positive rescaling $\tau \mapsto c \tau,$ which introduces only a state-independent additive ambiguity $S_{\tau} \mapsto S_{\tau} + \log{c}.$
In type II factors, therefore, the entropy difference between two density matrices can be unambiguously defined.
There is a natural procedure for assigning a density matrix in $\A_{\text{s.c.}}$ to any pure state in $\H_{\text{s.c}},$ and when this is done for pure states describing black holes with slowly varying wavefunctions for their gravitational fields, the resulting entropy difference is consistent with the expected entropy difference of the Bekenstein-Hawking formula \eqref{eq:Bekenstein-Hawking} \cite{Chandrasekaran:deSitter, Chandrasekaran:microcanonical, Jensen:general-subregions, Kudler-Flam:black-holes, Kudler-Flam:approximation}.

While the von Neumann algebra technology permits one to define and compute unambiguous entropy differences between perturbatively similar black holes, it is not totally clear from this analysis alone that the semiclassical entropy differences actually capture the entropy differences in the quantum theory.
In other words, if we take a pair of fully quantum black hole states that have finite entropy difference in the semiclassical limit, we would certainly like it to be the case that the limit of these fundamental entropy differences agrees with the ``bottom-up'' entropy difference computed using von Neumann algebras.
This can be seen in certain settings in $\mathcal{N}=4$ super Yang-Mills theory where the black hole wavefunctions at finite $G_N \hbar$ are holographically dual to precisely specified quantum field theory wavefunctions at finite gauge rank $N$ \cite{Witten:crossed-product, Chandrasekaran:microcanonical}; however, a general understanding of the emergence of type II entropy differences from fully quantum entropy differences remains elusive.\footnote{See however \cite{Soni:type-I, Akers:one-shot}.}

One step toward verifying the physical relevance of type II entropy differences was provided by myself and Akers in \cite{Akers:state-counting}.
There, we showed that in type II factors, without any reference to a limiting family of finite-dimensional quantum systems, the renormalized entropy difference of density matrices has a physical interpretation as a relative count of states.
We considered microcanonical density matrices for the black hole mass, which are simply projection operators in $\A_{\text{s.c.}}$ with finite renormalized trace.
The entropy difference between two such states $\rho_1$ and $\rho_2$ is
\begin{equation}
	\Delta S = \log \frac{\tau(\rho_1)}{\tau(\rho_2)}.
\end{equation}
In \cite{Akers:state-counting}, we showed that the ratio of $\tau(\rho_1)$ to $\tau(\rho_2)$ represents the additional (finite) number of quantum degrees of freedom that are needed to represent the state $\rho_1$ provided that one already has access to the state $\rho_2.$
Within the type II setting all density matrices have infinite rank; it is the relative index of this infinity that is captured by $\Delta S.$

Concretely, let $\H_j$ be the microcanonical support of $\rho_j.$
In \cite{Akers:state-counting}, using mathematical results from \cite{rings-1, rings-2, rings-3, rings-4}, it was demonstrated that one has $\Delta S = 0$ if and only if there is a partial isometry $V \in \A_{\text{s.c.}}$ satisfying
\begin{align}
	V \H_2
		& = \H_1, \\
	V^{\dagger} \H_1
		& = \H_2. 
\end{align}
Furthermore, one has $e^{\Delta S} = n \in \mathbb{Z}$ if and only if there is a pairwise-orthogonal family of partial isometries $V_j \in \A_{\text{s.c.}}$ satisfying
\begin{align}
	\sum_{j=1}^{n} (V_j \otimes \bra{j}) (\H_2 \otimes \comps^n)
		& = \H_1, \\
	\sum_{j=1}^{n} (V_j^{\dagger} \otimes \ket{j}) \H_1
		& = \H_2 \otimes \comps^n.
\end{align}
In \cite{Akers:state-counting}, we interpreted this as the statement that one can get from $\H_2$ to $\H_1$ using operations in the algebra if and only if one possesses an additional $e^{\Delta S}$ degrees of freedom.

The necessity of restricting to partial isometries contained in the algebra $\A_{\text{s.c}}$ raises some questions.
Since $\H_1$ and $\H_2$ are two infinite-dimensional subspaces of the same Hilbert space $\H_{\text{s.c.}}$, there always exists \textit{some} partial isometry on $\H_{\text{s.c.}}$ that maps $\H_1$ to $\H_2$; it is only when we restrict ourselves to work with isometries in the semiclassical algebra $\A_{\text{s.c}}$ that the entropy difference becomes constraining.

The purpose of this short paper is to shine a light on some of the mysteries discussed so far using an analogy with the emergence of classical phase space from a quantum mechanical system in the limit $\hbar \to 0.$
Classical states are described by probability distributions on phase space.
The preferred quantum measure on phase space is $dx dp/\hbar$, but this blows up in the classical limit, and must be ``renormalized'' to a scale-ambiguous measure $c\, dx dp.$
This makes entropy differences well defined in the classical limit, even though absolute entropies are infinite.
In this setting, one can explicitly find families of quantum states for which the classical limit of the quantum entropy difference matches the classically computed value.
The point of this paper is to show that in such a setting, there is an avatar of the von Neumann algebra restriction: the supports of two topologically equivalent uniform distributions on phase space can always be mapped into each other by some diffeomorphism, but to map them to one another using a \textit{symplectomorphism}, the phase space volumes must be the same.
The restriction that the map be a symplectomorphism is the analogue of the restriction that one must map microcanonical supports using partial isometries contained in the semiclassical algebra.
What is interesting about this setting is that we can see explicitly how the symplectomorphism restriction comes from the underlying quantum theory --- a unitary transformation of the quantum Hilbert space becomes, in the classical limit, a symplectomorphism of phase space.
This gives some indication of how the partial-isometry restriction of \cite{Akers:state-counting} should arise from the underlying theory of quantum gravity --- the partial isometries in the type II algebra that map between microcanonical supports are the semiclassical limits of unitary transformations on the quantum gravity Hilbert space; if we had not made the restriction that these partial isometries lie in the algebra, then we would not expect to be able to reproduce them in the underlying quantum gravitational theory.

In the next (and only) section, I describe two simple examples in which all of these ideas can be realized explicitly.

The first example is the theory of a free particle on a circle, and the relevant microcanonical states are supported in a momentum window that can be freely chosen.
At finite $\hbar$, these states are described by a fairly complicated Wigner function on the $(x,p)$ phase space; in the classical limit, these become uniform probability distributions on the portions of phase space bounded by the relevant momenta.
It is shown that the classical entropy differences of these probability distributions agree with the classical limits of the quantum entropy differences in the underlying theory, and that two such distributions have the same classical entropy if and only if they are related by a symplectomorphism.
Moreover, it is shown how the relevant symplectomorphisms arise as the classical limits of unitary transformations in the underlying quantum theory.

The second example shows similar results for microcanonical energy states of the quantum harmonic oscillator, in which case both the approach to classicality and the symplectomorphisms used in the classical limit are less trivial than in the case of the free particle on a circle.

\section{The phase space analogy}

\subsection{Wigner functions} \label{subsec:wigner}

Consider a theory of a single particle.
Classically, this is described by the phase space $(x,p)$ with the symplectic form $dx \wedge dp$ and a Hamiltonian functional $H(x,p)$.
Quantumly, the theory is described by the Hilbert space $\H = L^2(\reals),$ with operators $\hat{x}$ and $\hat{p} = - i \hbar \frac{\del}{\del x},$ together with a Hamiltonian operator $\hat{H}.$
To study the classical limit of a quantum system, it is helpful to rewrite the standard Hilbert space language in terms that explicitly reference phase space, even at nonzero $\hbar.$
These terms are provided by the Wigner transformation \cite{Wigner:original}.

The Wigner transformation provides a way to rewrite both generic density matrices $\rho$ and generic operators $\mathcal{O}$ as number-valued functions on phase space, so that the expectation value of an operator with respect to a given density matrix is just computed by integration of the corresponding functions.
The Wigner function of a density matrix $\rho$ is
\begin{equation}
	\rho_{W}(x, p) \equiv \frac{1}{2 \pi} \int dy \left\langle x - \frac{y}{2} \right| \rho \left| x + \frac{y}{2} \right\rangle e^{i p y/\hbar},
\end{equation}
and the Wigner function of an observable $\mathcal{O}$ is the same but without the prefactor:
\begin{equation}
	\mathcal{O}_{W}(x, p) \equiv \int dy \left\langle x - \frac{y}{2} \right| \mathcal{O} \left| x + \frac{y}{2} \right\rangle e^{i p y/\hbar}.
\end{equation}
From these formulas it is straightforward to verify the identity
\begin{equation}
	\tr(\rho \mathcal{O}) = \int \frac{dx\, dp\,}{\hbar} \rho_{W}(x, p) \mathcal{O}_{W}(x, p).
\end{equation}
In other words, quantum expectation values are represented as integrals of the corresponding Wigner functions with respect to the dimensionless phase space measure $dx dp/\hbar$.

All of the information about a quantum state $\rho$ is therefore contained in the corresponding Wigner function $\rho_W(x, p).$
This is generically not a probability distribution; it is always real-valued, and always integrates to one with respect to the measure $dx dp/\hbar$, but it can take negative values.
Given a family of quantum states parametrized by $\hbar,$ one can study how $\rho_W(x,p)$ behaves as a function of $\hbar$; if the family of quantum states has a good classical limit, then $\rho_W(x, p) / \hbar$ will become a classical probability distribution at $\hbar = 0$ with respect to the $\hbar$-renormalized measure $dx dp$; i.e., it will become a state in classical statistical mechanics.\footnote{As an instructive example, one can easily verify that for the ground state of the simple harmonic oscillator, the Wigner function is
\begin{equation}
	\rho_W(x,p) = \frac{1}{\pi} e^{- \frac{p^2+m^2 \omega^2 x^2}{\hbar m \omega}},
\end{equation}
from which one can see that in the classical limit we have $\rho_{W}(x,p)/\hbar \to \delta(x)\delta(p)$.
This is the expected classical behavior of the quantum ground state: it becomes the classical trajectory that sits at the origin of phase space forever.}
Moreover, one can verify that if $\mathcal{O}$ is a polynomial in $\hat{x}$ and $\hat{p}$, then in the $\hbar \to 0$ limit the Wigner function $O_W(x,p)$ becomes the corresponding polynomial in $x$ and $p$.\footnote{It suffices to check this for monomials of the form $\hat{x}^m \hat{p}^n,$ since all other polynomials can be expressed as linear combinations of these up to commutator terms that vanish in the $\hbar \to 0$ limit.
For such a monomial, the Wigner function is easily computed to be
\begin{equation}
	 e^{i \frac{2 x p}{\hbar}} \left(\frac{i \hbar}{2} \frac{\del}{\del p} \right)^m (p^n e^{- i \frac{2 x p}{\hbar}}),
\end{equation}
which in the $\hbar \to 0$ limit goes to $x^m p^n.$}

\subsection{Example 1: Free particle on a circle}

In an effort to analogize the transition from quantum black holes to semiclassical ones, we now turn our attention to the simplest quantum system with a preferred discretuum of states that becomes a continuum in the classical limit: the theory of a free particle on a circle of radius 1.
We can express the configurations of this theory in the standard $(x, p)$ phase space by imposing periodic boundary conditions at $x=0$ and $x=2 \pi,$ and demanding that all wavefunctions vanish outside of the interval $x \in [0, 2 \pi].$
The momentum eigenstates are labeled by integers $n$, and have wavefunctions
\begin{equation}
	\langle x | n \rangle = \frac{1}{\sqrt{2 \pi}} e^{i n x} \Theta(x \in [0, 2 \pi]).
\end{equation}
The Wigner functions of the corresponding density matrices $\rho^{(n)} = \ketbra{n}$ are easy to compute; they are given by
\begin{equation} \label{eq:Wigner-n}
	\frac{\rho^{(n)}_{W}(x, p)}{\hbar}
		= \frac{1}{2 \pi^2 (\hbar n-p)} \sin\left( \frac{2(\hbar n-p)(\pi - |\pi - x|)}{\hbar} \right) \Theta(x \in [0, 2 \pi]).
\end{equation}

To take a classical limit, we must specify an $\hbar$-dependent family of quantum states.
The simplest thing to do is to fix $n,$ so that we are always considering the $n$-th momentum eigenvalue as $\hbar$ goes to zero.
Because the momentum of the state $\rho^{(n)}$ is $\hbar n,$ any fixed-$n$ family of states should condense to the zero-momentum classical state; naively, we expect this to be the probability distribution $\delta(p)/2\pi$ in phase space, which is uniformly distributed in position but fixed to have zero momentum.
Indeed this is the case.
The easiest way to see this is to Fourier transform equation \eqref{eq:Wigner-n} with respect to $p,$ which gives
\begin{equation}
	\int dp\, \frac{\rho^{(n)}_{W}(x, p)}{\hbar} e^{-i p \nu}
	= \frac{1}{2 \pi} e^{-i \hbar n \nu} \Theta\left( |\nu| \leq \frac{2 (\pi - |\pi - x|)}{\hbar}\right).
\end{equation}
At fixed $n,$ the limit $\hbar \to 0$ results in the constant function $1/2\pi,$ and the Fourier transform can then be inverted to give $\delta(p)/2\pi,$ as desired.

If, by contrast, we fix some momentum $p_0$ and consider the family of states with $n \approx p_0/\hbar$, then we expect the Wigner functions of the corresponding states to approach the classical probability distribution $\delta(p-p_0)/2 \pi.$
This is easily shown from the same method as in the preceding paragraph; we have
\begin{equation} \label{eq:second-example}
	\int dp\, \frac{\rho^{(p_0/\hbar)}_{W}(x, p)}{\hbar} e^{-i p \nu}
	= \frac{1}{2 \pi} e^{-i p_0 \nu} \Theta\left( |\nu| \leq \frac{2 (\pi - |\pi - x|)}{\hbar}\right),
\end{equation}
which in the $\hbar \to 0$ limit goes to the function $e^{-i p_0 \nu}/ 2 \pi$; again inverting the Fourier transform, we obtain the expected probability distribution $\delta(p-p_0)/2\pi.$

Now we turn our attention to the microcanonical density matrices in the quantum theory, as well as their classical avatars.
In the $\hbar \to 0$ limit of such states, we expect to recover some qualitative features of the $G_N \hbar \to 0$ limit of microcanonical black holes.
Fixing an interval $[p_-, p_+]$ of the real line, we can consider the microcanonical density matrix that is maximally mixed over the momentum eigenstates with momenta contained in this interval.
Up to small approximations due to the discreteness of $n,$ this density matrix looks like
\begin{equation}
	\rho^{[p_-, p_+]}
		= \frac{\hbar}{p_+ - p_-} \sum_{n=p_-/\hbar}^{p_+/\hbar} \rho^{(n)}.
\end{equation}
Because the momenta are fixed, we are studying something like the example from equation \eqref{eq:second-example} above; but because we are studying a mixture over a number of states that scales like $1/\hbar,$ there is an additional factor of $\hbar$ that affects the details of the classical limit.
Again we study the Fourier transform, which gives
\begin{equation}
	\int dp\, \frac{\rho^{[p_-, p_+]}_{W}(x, p)}{\hbar} e^{-i p \nu}
	= \frac{1}{p_+ - p_-} \frac{1}{2 \pi} \Theta\left( |\nu| \leq \frac{2 (\pi - |\pi - x|)}{\hbar}\right) \sum_{n=p_-/\hbar}^{p_+/\hbar} \hbar e^{-i \hbar n \nu}.
\end{equation}
In the $\hbar \to 0$ limit the Heaviside theta goes away as in the previous examples; furthermore, the sum is approximated by an integral:
\begin{equation}
	\lim_{\hbar \to 0} \int dp\,\frac{\rho^{[p_-, p_+]}_{W}(x, p)}{\hbar} e^{-i p \nu}
	= \frac{1}{p_+ - p_-} \frac{1}{2 \pi} \int_{p_-}^{p_+} dp\, e^{-i p \nu}.
\end{equation}
Inverting the Fourier transform gives the expected result
\begin{equation} \label{eq:microcanonical-classical-uniform}
	\lim_{\hbar \to 0} \frac{\rho^{[p_-, p_+]}_{W}(x, p)}{\hbar}
	= \frac{1}{2 \pi(p_+ - p_-)} \Theta(p \in [p_-, p_+]),
\end{equation}
which is just the uniform distribution over the classical momentum range in which the quantum state was supported.

We are now in a position to see the first aspect of this analogy that mimics a putative property of the transition from quantum to semiclassical black holes: an exact matching between entropy differences computed in the classical theory, and the finite limits of entropy differences coming from the underlying quantum theory.
This is almost trivial: the microcanonical state supported on the momentum range $[p_-, p_+]$ has entropy roughly $\log((p_+ - p_-)/\hbar),$ up to small errors coming from the discreteness of $n.$ 
This diverges in the $\hbar \to 0$ limit, but given another interval $[p'_-, p'_+],$ the entropy difference
\begin{equation} \label{eq:quantum-entropy-limit}
	\Delta S = \log\left(\frac{p_+ - p_-}{p'_+ - p'_-}\right) + \text{(errors vanishing in $\hbar$)}
\end{equation}
is finite in the $\hbar \to 0$ limit.
The classical entropy of a probability distribution $P$ is defined by
\begin{equation}
	S = - \int dx dp\, P(x, p) \log P(x, p),
\end{equation}
up to an inherent additive ambiguity inherited from the scaling ambiguity in the measure $dx dp.$
For two uniform distributions, the classical entropy difference is the logarithm of the ratio of the phase space volumes.
For the uniform distributions that arise via equation \eqref{eq:microcanonical-classical-uniform}, the ratio of these volumes clearly reproduces the answer \eqref{eq:quantum-entropy-limit} obtained as a limit of the quantum theory.

Thus far we have given an explicit and elementary demonstration of a setting where a family of microcanonical quantum states has finite entropy differences in the classical limit, and for which the limiting entropy differences are correctly computed by the classical theory.
Now we will aim to understand, in this explicit setting, what is the appropriate analogy for the results of \cite{Akers:state-counting}.
In other words, we will formulate an information-theoretic description of the classical entropy differences for the phase space probability distributions, then show how this is inherited from the information-theoretic properties of the underlying quantum entropy.

Let us now suppose we have two uniform probability distributions on microcanonical portions of phase space, $P$ and $P',$ respectively supported on $[p_-, p_+]$ and $[p'_-, p'_+].$
The normalization of these probability distributions is subject to a scaling ambiguity in the classical theory, inherited from the scaling ambiguity of the dimensionful measure $dx dp.$
We will fix the measure $dx dp$ and normalize the probability distributions with respect to this choice; this is just like fixing a convention within the scale-ambiguous family of renormalized traces on a type II factor.
Concretely, we have
\begin{equation}
	P(x,p)
		= \frac{1}{2 \pi(p_+ - p_-)} \Theta(p \in [p_-, p_+]),
\end{equation}
and similarly for $P'.$
We would like to say that if $P$ and $P'$ have the same entropy, then they have the ``same number of classical states.''
It is hard to formulate this precisely without reference to the underlying quantum theory, because there is a continuous infinity of classical states.
What is true is that the supports of $P$ and $P'$ have the same phase space volume, and that this statement is independent of the scaling ambiguity in $dx dp.$
One way of rephrasing this is that the supports of $P$ and $P'$ are related by a translation in the $p$ direction, and this translation is a symmetry of the underlying symplectic structure; i.e., it is a symplectomorphism.
Because symplectomorphisms are volume-preserving, we also know that if the phase space volumes are different, then there are no symplectomorphisms relating the supports.

Already in this simple example we see an analogue of the type-II-algebraic results from \cite{Akers:state-counting}.
We have a preferred class of transformations acting on the physical space of states --- these are the symplectomorphisms on phase space --- and two of our microcanonical states have the same classical entropy if and only if they are related by one of these maps.
It remains only to understand \textit{why} we should impose the restriction that one must act on phase space with symplectomorphisms; in other words, we wish to understand how this restriction arises from the underlying quantum theory.

Given two families of microcanonical states, one supported on $[p_-, p_+]$ and one supported on $[p'_-, p'_+],$ and supposing that these two intervals have the same length, then there is a natural unitary that relates the density matrices up to discreteness errors.
This is the unitary map
\begin{equation} \label{eq:particle-unitary-family}
	U \ket{n} = \left| n + \delta(\hbar)\right\rangle,
\end{equation}
with
\begin{equation}
	\delta(\hbar) = \left[\frac{p_+ - p_+'}{\hbar} \right],
\end{equation}
and where $[x]$ denotes the nearest integer to the real number $x.$
By acting with $U$ on a generic operator $\mathcal{O}$, we obtain an action of $U$ on arbitrary Wigner functions.
To see how this acts on phase space in the classical limit, we just need to find operators whose Wigner functions in the classical limit are the phase space coordinates $x$ and $p$, and see how $U$ acts on these operators.
On the full $(x, p)$ plane, we already saw in section \ref{subsec:wigner} that the Wigner function of $\hat{x}$ is $x,$ and that the Wigner function of $\hat{p}$ is $p.$
On the restricted, periodic phase space $x \in [0, 2\pi],$ it is still true that the Wigner function of $\hat{x}$ is $x,$ but due to the restricted range of integration, the Wigner function of $\hat{p}$ is
\begin{align}
	\begin{split}
	\hat{p}_{W}(x,p)
		& = \int_{-2(\pi-|\pi-x|)}^{2(\pi-|\pi-x|)} dy\, \langle x-y/2|\hat{p}|x+y/2\rangle \\
		& = \int_{-\infty}^{\infty} dy\, \int_{-\infty}^{\infty} dq \frac{q e^{i(p-q)y/\hbar}}{2 \pi \hbar} \Theta\left(|y| \leq 2(\pi - |\pi - x|) \right).
	\end{split}
\end{align}
This is not equal to $p,$ but one can substitute $\nu = y/\hbar$ and find
\begin{align}
	\begin{split}
		\hat{p}_{W}(x,p)
		& = \int_{-\infty}^{\infty} d\nu\, \int_{-\infty}^{\infty} dq \frac{q e^{i(p-q)\nu}}{2 \pi} \Theta\left(|\nu| \leq \frac{2(\pi - |\pi - x|)}{\hbar} \right)
	\end{split}
\end{align}
Taking the $\hbar \to 0$ limit before integrating, we find that this converges to $p$ in the limit $\hbar \to 0.$
It is clear that the family of unitaries in equation \eqref{eq:particle-unitary-family} maps $\hat{p}$ to $\hat{p} + \hbar \delta(\hbar),$ which in the classical limit is $p \mapsto p + (p_+ - p_+').$
It is also true, at any value of $\hbar$, that $U$ maps $\hat{x}$ to itself --- to see this, we compute
\begin{align}
	\begin{split}
		U \hat{x} U^{\dagger}
			& = \sum_{m,n} U \ketbra{m} \hat{x} \ketbra{n} U^{\dagger} \\
			& = \int dx\, \sum_{m, n} \frac{x e^{-i (m-n) x}}{2 \pi} \ketbra{m+\delta}{n+\delta}.
	\end{split}
\end{align}
Relabeling the sum as $m \mapsto m-\delta, n \mapsto n - \delta$ gives
\begin{align}
	\begin{split}
		U \hat{x} U^{\dagger}
		& = \int dx\, \sum_{m, n} \frac{x e^{-i (m-n) x}}{2 \pi} \ketbra{m}{n} \\
		& = \hat{x}.
	\end{split}
\end{align}
So in total, the classical limit of $U$ is the translation $x \mapsto x, p \mapsto p + (p_+ - p_+').$

To summarize, the technology of Wigner functions made it easy to see:
\begin{itemize}
	\item that microcanonical momentum states for the free particle on a circle become, in the classical limit, uniform distributions;
	\item that the classical entropy differences of these uniform distributions are the same as the classical limits of the corresponding quantum entropy differences;
	\item that two of these classical probability distributions have the same entropy if and only if they are related by a symplectomorphism; and finally
	\item that these classical symplectomorphisms arise explicitly as the classical limit of unitaries that relate the microcanonical states in the underlying quantum description.
\end{itemize} 
The first and third of these are directly analogous to the results of \cite{Akers:state-counting} for semiclassical black holes; the second and fourth are explicitly realized in the present setting, and analogues are expected to hold when the construction of \cite{Akers:state-counting} is embedded into a full theory of quantum gravity.

To conclude, we will now see that the same basic structure appears for microcanonical energy states in the simple harmonic oscillator, even though the structure of the symplectomorphisms is more complicated.

\subsection{Example 2: The quantum harmonic oscillator} \label{subsec:SHO}

The previous example of the free particle on a circle contained all of the conceptual pieces we aimed to understand --- symplectomorphism equivalence as a method for comparing classical state-counts, and the relation between classical symplectomorphisms and quantum unitaries --- but one might wonder, because the relevant symplectomorphisms were just translations in the $p$ direction, if these features had more to do with the extreme simplicity of the example than with fundamental properties of the classical limits of quantum systems.
To alleviate this concern, here we show analogous results for microcanonical energy states of the quantum harmonic oscillator, for which the relevant symplectomorphisms are less trivial; in fact they are non-invertible, and they arise as the classical limit of quantum isometries, not quantum unitaries.

The $n$-th energy eigenstate of the quantum harmonic oscillator, written as $|n\rangle,$ has position-space wavefunction
\begin{equation}
	\langle x | n \rangle
		= \frac{1}{\sqrt{2^n n!}} \left(\frac{m \omega}{\pi \hbar} \right)^{1/4}
			e^{-\frac{m \omega x^2}{2 \hbar}} H_n \left( \sqrt{\frac{m \omega}{\hbar}} x\right),
\end{equation}
where $H_n$ is the $n$-th Hermite polynomial defined by the generating function
\begin{equation} \label{eq:Hermite-generating-n}
	H_n(z) = \left.\frac{\del^n}{\del t^n} \left( e^{2 z t-t^2} \right)\right|_{t=0}.
\end{equation}
To find the Wigner function of the $n$-th energy eigenstate, we must evaluate the integral
\begin{align}
	\begin{split}
	\rho^{(n)}_W(x,p)
		& = \frac{1}{2\pi} \int dy\, \left\langle x - \frac{y}{2} \right| \rho^{(n)} \left| x + \frac{y}{2} \right\rangle e^{ipy/\hbar}.
	\end{split}
\end{align}
To save some space in the intermediary calculations, we will write $\alpha = \sqrt{m \omega/\hbar}$.
The integral we want to evaluate is
\begin{equation}
	\rho^{(n)}_W(x,p)
		= \frac{1}{2\pi} \frac{\alpha e^{- \alpha^2 x^2}}{2^n n! \sqrt{\pi}} \int dy\, e^{i p y/\hbar}
		e^{- \alpha^2 y^2/4}
		H_n \left( \alpha \left(x - \frac{y}{2}\right) \right)
		H_n \left( \alpha \left(x+\frac{y}{2}\right) \right).
\end{equation}
If we substitute for each $H_n$ the generating functional expression \eqref{eq:Hermite-generating-n}, and pull the derivatives outside of the integral, we are left with the expression
\begin{equation}
	\rho^{(n)}_W(x,p)
	= \frac{1}{2\pi} \frac{\alpha e^{- \alpha^2 x^2}}{2^n n! \sqrt{\pi}} \left. \frac{\del^{2n}}{\del t_1^n \del t_2^n} \int dy\, e^{i p y/\hbar}
	e^{- \alpha^2 y^2/4}
	e^{2 \alpha \left(x - \frac{y}{2}\right) t1 - t1^2} e^{2 \alpha \left(x + \frac{y}{2}\right) t_2 - t_2^2} \right|_{t_1=t_2=0}.
\end{equation}
The integral is Gaussian and can be evaluated explicitly; the answer is 
\begin{equation}
	\rho^{(n)}_W(x,p)
		= \frac{1}{\pi} \frac{e^{- \alpha^2 x^2-\frac{p^2}{\hbar^2 \alpha^2}}}{2^n n!} \left. \frac{\del^{2n}}{\del t_1^n \del t_2^n}
		\left( e^{2 t_1 \left(x \alpha - i \frac{p}{\hbar \alpha}\right)}
		e^{2 t_2 \left(x \alpha + i \frac{p}{\hbar \alpha}\right)}
		e^{- 2 t_1 t_2} \right)
		\right|_{t_1=t_2=0}.
\end{equation}
The three exponentials in the parentheses can be expanded as power series, from which the terms proportional to $t_1^n t_2^n$ can be easily extracted; from this, one finds the expression
\begin{equation}
	\rho^{(n)}_W(x,p)
	= \frac{(-1)^n}{\pi} e^{- \left(\frac{p^2}{\hbar^2 \alpha^2} + \alpha^2 x^2 \right)}
	\sum_{j=0}^n {n\choose j} \frac{(-1)^j}{j!} \left( \frac{2 p^2}{\hbar^2 \alpha^2} + 2 \alpha^2 x^2 \right)^j.
\end{equation}
This is an explicit expression for the Laguerre polynomial $L_n,$ defined by
\begin{equation}
	L_n(z)
		= \frac{1}{n!} e^{z} \frac{\del^n}{\del z^n} (z^n e^{-z})
		= \sum_{j=0}^n {n \choose j} \frac{(-1)^j}{j!} z^j. 
\end{equation}
This gives a concise expression for the Wigner function in terms of the classical Hamiltonian $H(x, p)$ as
\begin{equation} \label{eq:SHO-Wigner}
	\rho^{(n)}_W(x,p)
	= \frac{(-1)^n}{\pi} e^{- 2 H(x,p)/\hbar \omega}
	L_n\left(\frac{4 H(x,p)}{\hbar \omega}\right).
\end{equation}
This elementary computation matches the known answer for the Wigner functions of the energy eigenstates of the simple harmonic oscillator, which were first computed in \cite{Groenewold:thesis} in a different form.

As in the previous subsection, we may now examine the classical limits of various families of states.
The simplest case is one where we fix $n$.
This family of states has energy $\hbar \omega(n+1/2)$, so in the $\hbar \to 0$ limit we expect to localize on the unique zero-energy classical configuration, i.e., we expect to have $\rho_{W}^{(n)}(x, p)/\hbar \to \delta(x) \delta(p).$
Indeed, the Fourier transform of this expression in $p$ is just a slight rewriting of the original integrand for the Wigner function:
\begin{equation}
	\int dp\, \frac{\rho_W^{(n)}(x,p)}{\hbar} e^{-i p \nu}
		= \frac{\sqrt{\frac{m \omega}{\pi \hbar}}}{2^n n!} e^{-\frac{m \omega (4 x^2 + \hbar^2 \nu^2)}{4 \hbar}}
		H_n \left( \sqrt{\frac{m \omega}{\hbar}} \left( x-\hbar \frac{\nu}{2}\right)\right)
		H_n \left( \sqrt{\frac{m \omega}{\hbar}} \left( x+\hbar \frac{\nu}{2}\right)\right).
\end{equation}
In this expression the $\hbar \to 0$ limit kills all $\nu$-dependence, leaving
\begin{equation}
	\lim_{\hbar \to 0} \int dp\, \frac{\rho_W^{(n)}(x,p)}{\hbar} e^{-i p \nu}
	= \lim_{\hbar \to 0} \frac{1}{2^n n!} \sqrt{\frac{m \omega}{\pi \hbar}} e^{-\frac{m \omega x^2}{\hbar}}
	H_n \left( \sqrt{\frac{m \omega}{\hbar}} x \right)^2.
\end{equation}
The Fourier transform of this with respect to $x$ is easily computed using the generating functionals for the Hermite polynomial, as in the preceding paragraph, and the answer is
\begin{equation}
	\lim_{\hbar \to 0} \int dp\, \frac{\rho_W^{(n)}(x,p)}{\hbar} e^{-i p \nu - i x \mu}
	= \lim_{\hbar \to 0} e^{- \hbar \mu^2/4 m \omega} L_n \left( \frac{\hbar \mu^2}{m \omega}\right)
	= 1.
\end{equation}
Inverting the Fourier transform gives the expected answer $\delta(x) \delta(p).$

Next we turn our attention to the case where we fix the energy; from the relation $E = \hbar \omega (n + 1/2),$ we fix some energy $E_0$ and consider the $\hbar$-dependent family of energy eigenstates with $n = E_0/\hbar \omega - 1/2.$
Our expectation is that in the classical limit $\hbar \to 0,$ the Wigner function should localize to a delta distribution on the ellipse in phase space defined by $H(x,p) = E_0.$
Showing this entails taking the $\hbar \to 0$ limit (in the sense of distributions) of $\rho_W^{(E_0/\hbar \omega)}(x,p)/\hbar$.
To do this, we substitute $x = X/\sqrt{m}\omega$ and $p=\sqrt{m}P,$ then $X^2 + P^2 = R^2,$ and write
\begin{align}
	\begin{split}
	\frac{\rho_W^{(n)}(x,p)}{\hbar} dx\, dp
		& = \frac{(-1)^n}{\pi \hbar} e^{-2H(x,p)/\hbar \omega} L_n\left(\frac{4 H(x,p)}{\hbar \omega} \right) \\
		& = \frac{(-1)^n}{\pi \hbar \omega} e^{- (P^2+X^2)/\hbar \omega}
			L_n \left( 2 \frac{P^2 + X^2}{\hbar \omega} \right) dX dP \\
		& = \frac{(-1)^n}{\pi \hbar \omega} e^{- R^2/\hbar \omega} L_n\left( \frac{2 R^2}{\hbar \omega} \right) R dR d\theta.
	\end{split}
\end{align}
In the limit $\hbar \to 0$ with $n = E_0/\hbar\omega - 1/2,$ we want this to localize to
\begin{equation}
	\frac{1}{2\pi} \delta\left(R - \sqrt{2 E_0}\right) dR d\theta.
\end{equation}
This can be done using the asymptotic limits for the Laguerre polynomials from \cite{erdelyi1960asymptotic, frenzen1988uniform, temme1990asymptotic}; see appendix \ref{app:asymptotics} for an explicit demonstration. 

Now, as we did in the previous subsection for the free particle on a circle, we turn our attention to microcanonical density matrices and their classical limits.
We fix an interval $[E_-, E_+]$ in the positive real axis, and consider the microcanonical density matrix that is maximally mixed over the energy eigenstates in this window.
Up to small approximation errors due to the discreteness of $n,$ this density matrix is
\begin{equation}
	\rho^{[E_-, E_+]}
		= \frac{\hbar \omega}{E_+ - E_-} \sum_{n=\frac{E_-}{\hbar \omega} - \frac{1}{2}}^{\frac{E_-}{\hbar \omega} - \frac{1}{2}} \rho^{(n)}.
\end{equation}
Using the same methods we applied in the case of the free particle on a circle, one can easily show that the $\hbar \to 0$ limit of the Wigner function gives the uniform distribution in the expected range of energies; in the above variables, this is
\begin{equation}
	\lim_{\hbar \to 0} \frac{\rho^{[E_-, E_+]}_{W}}{\hbar} dx dp
		= \frac{1}{2\pi (E_+ - E_-)} \int_{E-}^{E_+} dE\, \left( \delta(R - \sqrt{2 E}) dR d\theta \right).
\end{equation}
In the $(x,p)$ variables, this is the uniform distribution on the squashed annulus
\begin{equation}
	E_- \leq \sqrt{\frac{p^2}{2 m} + \frac{m \omega^2 x^2}{2}} \leq E_+.
\end{equation}
The phase space volume of such a region is $C (E_+ - E_-),$ where $C$ is a constant that depends on the chosen normalization of the measure on phase space.
The classical entropy difference of two such uniform distributions is therefore
\begin{equation}
	\Delta S = \log\left(\frac{E_+ - E_-}{E'_+ - E'_-}\right),
\end{equation}
which agrees with the $\hbar \to 0$ limit of the entropy differences of the microcanonical states in the underlying quantum theory.

Now let us turn to the question of how the classical entropy difference is interpreted in the language of symplectomorphisms.
When $\Delta S$ is zero --- i.e., we have $E_+ - E_- = E'_+ -  E'_-$ --- then there is a natural symplectomorphism relating the corresponding squashed annuli in phase space.
It is the transformation that is rotationally symmetric with respect to the $\theta$ coordinate used above, and that satisfies
\begin{equation}
	H(x,p) \mapsto H(x,p) + E_+ - E_+'.
\end{equation}
This is given in terms of $x$ and $p$ by
\begin{equation} \label{eq:SHO-symplectomorphism}
	\left( x, p \right)
		\mapsto \left( x \sqrt{\frac{H(x,p) + E_+ - E_+'}{H(x,p)}}\,\,,\,\, p \sqrt{\frac{H(x,p) + E_+ - E_+'}{H(x,p)}} \right)
\end{equation}
This is a nonlinear transformation of phase space, but from this formula it is easy to verify that it is a symplectomorphism --- though it is not invertible.
We aim to show that it arises as the classical limit of the family of isometries in the underlying quantum description given by
\begin{equation}
	V |n\rangle = |n + \delta(\hbar)\rangle
\end{equation}
with
\begin{equation}
	\delta(\hbar) = \left[ \frac{E_+ - E_+'}{\hbar \omega} \right], 
\end{equation}
where as in the previous subsection we have used the notation $[x]$ for the nearest integer to $x.$

The partial isometry $V$ clearly maps the Hamiltonian $\hat{H}$ to $\hat{H} + E_+ - E_+'$ up to discreteness errors.
Since the Wigner function of $\hat{H}$ goes to $H(x,p)$ in the classical limit, we know that the classical limit of $V$ maps $H(x,p)$ to $H(x,p) + E_+ - E_+'$; we must show that it is also angle-preserving.
To accomplish this, it suffices to find any observable that depends only on $x/p$ in the classical limit, and to study its action under $V$.
We choose the observable $(\hat{x}\, \hat{p} + \hat{p}\, \hat{x}) \hat{H}^{-1}$.
A straightforward calculation gives the matrix elements
\begin{equation}
	\langle j| (\hat{x}\, \hat{p} + \hat{p}\, \hat{x}) \hat{H}^{-1} |k\rangle
		= \frac{i \hbar}{E_k}\left( \sqrt{j(j-1)} \delta_{j,k+2} - \sqrt{k(k-1)} \delta_{j+2,k} \right)
\end{equation}
In the classical limit $\hbar \to 0,$ substituting energy variables for integer ones, we obtain
\begin{equation}
	(\hat{x}\, \hat{p} + \hat{p}\,\hat{x}) \hat{H}^{-1}
		\approx \int dE\, \frac{2 i}{\hbar \omega^2} \left( \ketbra{E}{E - 2 \hbar \omega} - \ketbra{E - 2 \hbar \omega}{E} \right).
\end{equation}
Consequently, when applying $V$, all that happens in the classical limit is that the range of integration is restricted to $E \geq E_+ - E_+'$.
This induces a map of the classical Wigner functions:
\begin{equation}
	V: \frac{2 x p}{H(x,p)} \mapsto \frac{2 x p}{H(x,p)} \Theta(H(x,p) \geq E_+ - E_+').
\end{equation}
Consequently, $V$ preserves the ratio $x/p$ in the classical limit and maps $H(x,p)$ to $H(x,p) + E_+ - E_+'$; this fixes it to be the symplectomorphism specified by equation \eqref{eq:SHO-symplectomorphism}.

\acknowledgments{This research was supported by the DOE Early Career Award DE-SC0021886,  the Packard Foundation Award in Quantum Black Holes and Quantum Computation, the DOE QuantISED DE-SC0020360 contract 578218, and by the Heising-Simons Foundation grant 2023-443.}

\appendix

\section{Classical limit of the harmonic oscillator Wigner function}
\label{app:asymptotics}

This supplement to section \ref{subsec:SHO} demonstrates the distributional identity
\begin{equation}
	\lim_{\hbar \to 0} \frac{(-1)^n}{\pi \hbar \omega} e^{- R^2/\hbar \omega} R\, L_n\left(\frac{2 R^2}{\hbar \omega} \right)
	= \frac{1}{2\pi} \delta(R - \sqrt{2 E_0}),
\end{equation}
with $n = E_0/\hbar \omega - 1/2.$
Substituting for $\hbar$ in terms of $n$, our aim is to show
\begin{equation} \label{eq:desired-distributional-limit}
	\lim_{n \to \infty} \frac{(-1)^n}{\pi} \frac{n + 1/2}{E_0} e^{- (n+1/2) R^2/E_0} R\, L_n\left(\frac{2 R^2}{E_0} \left(n + \frac{1}{2}\right) \right)
	= \frac{1}{2\pi} \delta(R - \sqrt{2 E_0}).
\end{equation}
Asymptotic expressions for the Laguerre polynomial in this regime are known thanks to \cite{erdelyi1960asymptotic, frenzen1988uniform, temme1990asymptotic}.
Translating from those works to the present setting, one writes the following expressions:
\begin{align}
	A(x)
		& = \frac{1}{2} \left(\sqrt{x - x^2} + \arcsin(\sqrt{x})\right), \\
	\alpha(x)
		& = \frac{\sqrt{A(x)}}{(x(1-x))^{1/4}}, \\
	B(x)
		& = i \left( \frac{3}{4} \left( \arccos(\sqrt{x}) - \sqrt{x-x^2} \right) \right)^{1/3}, \\
	\beta(x)
		& = \frac{\sqrt{2 |B(x)|}}{(x(1-x))^{1/4}}, \\
	C(x)
		& = \left( \frac{3}{4} \left( \sqrt{x^2-x} - \operatorname{arccosh}(\sqrt{x}) \right) \right)^{1/3} \\
	\gamma(x)
		& = \frac{\sqrt{2 |C(x)|}}{(x(x-1))^{1/4}}.
\end{align}
Then one finds the asymptotic expressions (for any $a,b < 1,$ and with $J_0$ the Bessel function and $\operatorname{Ai}$ the Airy function)
\begin{align} \label{eq:Laguerre-asymptotics}
	\begin{split}
		& e^{-(n+1/2) R^2/E_0} L_n\left(\frac{2 R^2}{E_0} \left(n + \frac{1}{2}\right)\right) \\
		& \qquad
		\approx_{n \to \infty} \begin{cases}
			J_0\left[4 \left(n + \frac{1}{2}\right) A\left( \frac{R^2}{2 E_0} \right) \right]
				& 0 \leq \frac{R^2}{2 E_0} \leq a < 1 \\
		\frac{(-1)^n}{(4n+2)^{1/3}} \beta\left(\frac{R^2}{2 E_0}\right) \operatorname{Ai}\left((4n+2)^{2/3} B\left(\frac{R^2}{2 E_0}\right)^2\right)
			& b \leq \frac{R^2}{2 E_0} \leq 1 \\
		\frac{(-1)^n}{(4n+2)^{1/3}} \gamma\left(\frac{R^2}{2 E_0}\right) \operatorname{Ai}\left((4n+2)^{2/3} C\left(\frac{R^2}{2 E_0}\right)^2\right)
		& 1 \leq \frac{R^2}{2 E_0} < \infty \\
		\end{cases}
	\end{split}
\end{align}
The first line can be used as an approximation on any interval of the form $[0, a]$ with $a < 1$; the second and third combined can be used as an approximation on any half-line of the form $[b, \infty)$ with $b < 1.$
From the third line it is straightforward to show that for any compact range of $R$ strictly bigger than $\sqrt{2 E_0}$, we have
\begin{equation}
	\lim_{n \to \infty} \frac{(-1)^n}{\pi} \frac{n+1/2}{E_0} R L_n \left( \frac{2 R^2}{E_0} \left(n + \frac{1}{2} \right)\right) \to 0,
\end{equation}
which follows from the exponential decay of the Airy function at large argument.
We wish to use the first line of equation \eqref{eq:Laguerre-asymptotics} to show a similar distributional identity for any compact range of $R$ strictly less than $2 E_0$.
To do this one uses the integral identity for the Bessel function,
\begin{equation}
	J_0(x) = \frac{1}{\pi} \text{Re}\left( \int_0^{\pi} d\tau e^{- i x \sin{\tau}}\right).
\end{equation}
Applying this to the first line in \eqref{eq:Laguerre-asymptotics}, one sees that the distributional limit of the LHS of \eqref{eq:desired-distributional-limit} is computed, in the regime where $R$ is strictly less than $\sqrt{2 E_0},$ by a saddle point analysis with respect to the phase $e^{i \# n A\left(\frac{R^2}{2 E_0}\right)}.$
But this phase has no saddles for $R < \sqrt{2 E_0}$, and the distributional limit is zero.

From the analysis so far we know that the distributional limit of the LHS of \eqref{eq:desired-distributional-limit} is supported entirely at $R = 2 E_0$; i.e., it is a finite sum of delta functions and derivatives of delta functions localized at this point.
Near $R=2 E_0$, we can expand $\beta, \gamma, B,$ and $C$, and use the second and third lines of equation \eqref{eq:Laguerre-asymptotics} to obtain the approximation
\begin{align}
	\begin{split}
	& \lim_{n \to \infty} \frac{(-1)^n}{\pi} \frac{n + 1/2}{E_0} e^{- (n+1/2) R^2/E_0} R\, L_n\left(\frac{2 R^2}{E_0} \left(n + \frac{1}{2}\right) \right) \\
	& \qquad = \frac{1}{2\pi} \lim_{n \to \infty} \frac{R}{E_0} \left(2n+1\right)^{2/3} \operatorname{Ai}\left(\sqrt{2} (2n+1)^{2/3} \frac{R - \sqrt{2 E_0}}{\sqrt{E_0}}\right)
	\end{split}
\end{align}
From the Fourier transform of the Airy function,
\begin{equation}
	\int dx\, \operatorname{Ai}(x) e^{- i \nu x} = e^{\frac{i}{3} \nu^3},
\end{equation}
one can see that the Fourier transform of $(2n+1)^{2/3}\operatorname{Ai}((2n+1)^{2/3} x)$ goes to one in the limit $n \to \infty$; hence we have $(2n+1)^{2/3}\operatorname{Ai}((2n+1)^{2/3} x) \to \delta(x),$ hence
\begin{align}
	\begin{split}
		& \lim_{n \to \infty} \frac{(-1)^n}{\pi} \frac{n + 1/2}{E_0} e^{- (n+1/2) R^2/E_0} R\, L_n\left(\frac{2 R^2}{E_0} \left(n + \frac{1}{2}\right) \right) \\
		& \qquad = \frac{1}{2\pi} \lim_{n \to \infty} \frac{R}{E_0} \delta\left( \sqrt{2} \frac{R - \sqrt{2 E_0}}{\sqrt{E_0}}\right) \\
		& \qquad = \frac{1}{2\pi} \delta(R - \sqrt{2 E_0}),
	\end{split}
\end{align}
as desired.

\bibliographystyle{JHEP}
\bibliography{bibliography}

\end{document}